\documentclass[%
reprint,
 superscriptaddress,
 amsmath,amssymb,
 aps,
 prl,
floatfix,
]{revtex4-1}%revtex4-2

\usepackage[dvipdfmx]{graphicx}% Include figure files
\usepackage{dcolumn}% Align table columns on decimal point
\usepackage{bm}% bold math
\usepackage{hyperref}% add hypertext capabilities
\usepackage[mathlines]{lineno}% Enable numbering of text and display math
\usepackage[T1]{fontenc}%for polish name

\newcommand{\RRO}{$R_2\mathrm{Ru}_2\mathrm{O}_7$}

\newcommand{\PCRO}{$(\mathrm{Pr}_{1-x}\mathrm{Ca}_x)_2\mathrm{Ru}_2\mathrm{O}_7$}
\newcommand{\GCRO}{$(\mathrm{Gd}_{1-x}\mathrm{Cd}_x)_2\mathrm{Ru}_2\mathrm{O}_7$}

\begin{document}

%\preprint{APS/123-QED}

\title{Fully filling-controlled pyrochlore ruthenates: emergent ferromagnetic-metal state and geometrical Hall effect }% Force line breaks with \\

\author{Ryoma Kaneko}
\affiliation{Department of Applied Physics, University of Tokyo, Tokyo 113-8656, Japan}%
\affiliation{RIKEN Center for Emergent Matter Science (CEMS), Wako 351-0198, Japan}%
\author{Kentaro Ueda}
\email[To whom correspondence should be addressed.\\]{ueda@ap.t.u-tokyo.ac.jp}%
%\altaffiliation[]{}%
%\homepage[]{}%
\affiliation{Department of Applied Physics, University of Tokyo, Tokyo 113-8656, Japan}%
\author{Shiro Sakai}
\affiliation{RIKEN Center for Emergent Matter Science (CEMS), Wako 351-0198, Japan}%
\author{Yusuke Nomura}
\affiliation{RIKEN Center for Emergent Matter Science (CEMS), Wako 351-0198, Japan}%
\author{Marie-Therese Huebsch}
\affiliation{RIKEN Center for Emergent Matter Science (CEMS), Wako 351-0198, Japan}%
\author{Ryotaro Arita}
\affiliation{Department of Applied Physics, University of Tokyo, Tokyo 113-8656, Japan}%
\affiliation{RIKEN Center for Emergent Matter Science (CEMS), Wako 351-0198, Japan}%
\author{Yoshinori Tokura}
\affiliation{Department of Applied Physics, University of Tokyo, Tokyo 113-8656, Japan}%
\affiliation{RIKEN Center for Emergent Matter Science (CEMS), Wako 351-0198, Japan}%
\affiliation{Tokyo College, University of Tokyo, Tokyo 113-8656, Japan}

\begin{abstract}
Carrier doping to the Mott insulator is essential to produce highly correlated metals with emergent properties. Pyrochlore ruthenates, $\mathrm{Pr}_2\mathrm{Ru}_2\mathrm{O}_7$ (Ru-$4d$ electron number, $n = 4$) and $\mathrm{Ca}_2\mathrm{Ru}_2\mathrm{O}_7$ ($n = 3$), are a Mott insulator and a magnetic bad metal, respectively, due to the strong electron correlation. We investigate magneto-transport properties of \PCRO\ in a whole band-filling range, $0 \leq x \leq 1$. With increasing hole-doping $x$, the system undergoes an insulator-metal transition. When $\mathrm{Ca}_2\mathrm{Ru}_2\mathrm{O}_7$ is doped with electrons ($0.5 < x < 0.9$), the enhanced coupling among Ru-$4d$ spins produces a ferromagnetic-metal phase with a large anomalous-Hall angle up to 2 \%. We discuss the electronic phase transitions in \PCRO\ in view of Hund's metal.
% \begin{description}
% Secondary publications and information retrieval purposes.
% You may use the \texttt{description} environment to structure your abstract;
% use the optional argument of the \verb+\item+ command to give the category of each item. 
% \end{description}
\end{abstract}
\maketitle
%
%\tableofcontents
%
\newpage
%\footnote{*corresponding author. email: ueda@ap.t.u-tokyo.ac.jp}
Mottness, a feature related to electron localization driven by on-site Coulomb repulsion $U$, is a key to emergent properties in strongly correlated electronic systems (SCES). The systematics of Mottness or related metal-insulator transition (MIT) has been intensively studied in $3d$-electron transition-metal oxides, where the bandwidth and band-filling can be systematically tuned via chemical doping or applying hydrostatic or chemical pressure~\cite{Imada1998}. Along with the variation of $U$ and band-filling, versatile magneto-electronic phenomena such as high-temperature superconductivity~\cite{Lee2004} and colossal magneto-resistance~\cite{Tokura2006} have been discovered in those systems, generally followed by MITs. Besides the Mottness characterized by $U$, recent studies also emphasize the importance of Hund's coupling $J_{\mathrm{H}}$~\cite{Georges2013,Deng2019}. Notably, $J_{\mathrm{H}}$ affects SCES in two different manners: (i) the partial screening of $U$ and the modification of the critical $U$ for the MIT, and (ii) the reduction of the coherent scale of quasiparticles via unscreened spin fluctuation. These effects of $J_{\mathrm{H}}$ may lead to a highly correlated non-Fermi liquid like state, so-called Hund's metal~\cite{Werner2008,Medici2011,Dang2015,Han2016,Nomura2015,Shinaoka2015,Nakatsuji2003}.
Those Hund's metal features are expected to dominate $4d$ and $5d$ electron systems, where the scales of $U$ and $J_{\mathrm{H}}$ are sufficiently close. These systems attract interest also in the light of topological properties derived from electronic bands reconstructed by strong spin-orbit coupling (SOC) $\lambda$~\cite{krempa2014correlated}. The interplay of $J_{\mathrm{H}}$ with $\lambda$ is nontrivial because $\lambda$ enhances the orbital mixing whereas $J_{\mathrm{H}}$ tends to decouple spin and orbital degrees of freedom~\cite{Kim2018}. Thus, the $4d$ and $5d$ electron systems attract current interest, where plural inter- and intra-atomic interactions cooperate or compete with each other.

In this context, \RRO\ ($R$ being a rare-earth element) provides an important arena for research. In \RRO, $R$ and Ru ions respectively form pyrochlore lattices composed of  corner-sharing tetrahedrons and displaced by a half unit cell from each other~\cite{Gardner2010}. In the ionic picture on the Mott insulator, local trigonal distortion around Ru$^{4+}$ splits the valence $t_{2g}$ orbitals into doubly-degenerated $e'_g$ and non-degenerated $a_{1g}$ orbitals; two of the four Ru-$4d$ electrons fully occupy the $a_{1g}$ orbital and the other two fill the half of the $e'_g$ orbtials with their spins aligned by Hund's-rule coupling [Fig.~1(b)]. As a result, \RRO\ has the $S = 1$ spin ordered antiferromagnetically at low temperatures while it is electrically insulator at all temperatures and for all $R$ (= Pr-Lu)~\cite{Taira1999,Ito2001}. The neutron diffraction studies suggest antiferromagnetic magnetic structures in \RRO, the details of which depend on the $R$-$4f$ moments with different magnetic anisotropies~\cite{Ito2001,Gardner2010,Ku2018,Gurgul2007,Kmiec2006, Taira2003}.
%Former studies predicted the non-collinear configuration of Ru moments~\cite{Ito2001,Gardner2010,Ku2018,Gurgul2007,Kmiec2006} while the collinear magnetic structure was predicted for $R$ = Er, whose $f$ moments are of Heisenberg $XY$ type~\cite{Taira2003}. 
The $R$ = Pr, Nd, or Ho compounds, whose $f$ moments have strong uniaxial anisotropy with respect to the local $\langle 111 \rangle$ axes, are especially important because these antiferromagnetic moments can order in a non-coplanar all-in-all-out (AIAO) structure [Fig.~1(a)]~\cite{Taira1999,Ito2001}. The magnetic field along the specific crystal axis can transform the AIAO to the 2-in-2-out or 3-in-1-out structure, which significantly affects the magnetic and electronic properties~\cite{Ueda2017}.

Trivalent $R$ ions can be replaced by divalent ions $A$ such as Ca and Cd, accompanied by the oxidization number increasing from Ru$^{4+}$ to Ru$^{5+}$. This reduces the number of Ru-electrons from 4 to 3. The spin state changes from $S$ = 1 to $S$ = 3/2 [Fig. 1(b)], and the upward shift of the oxygen $2p$ bands reduces $U$ and $J_{\mathrm{H}}$ by enhancing the screening effect. Contrary to \RRO, which is an antiferromagnetic Mott insulator, $\mathrm{Ca}_2\mathrm{Ru}_2\mathrm{O}_7$ and $\mathrm{Cd}_2\mathrm{Ru}_2\mathrm{O}_7$ show a bad metallic behavior with spin-glass transitions at low temperatures~\cite{Munenaka2006,Jiao2018}. Such contrasting magnetic and electronic characters between the end compounds, \RRO\ and $A_2\mathrm{Ru}_2\mathrm{O}_7$, imply novel phase transitions, as well as related magneto-electronic phenomena in the intermediate filling-controlled compounds $(R_{1-x}A_x)_2\mathrm{Ru}_2\mathrm{O}_7$.

A previous study~\cite{Kaneko2020} focused on the systematic change in charge dynamics of \RRO\ in the course of the bandwidth- and the filling-control MITs. In the relatively low hole-doped region, the MITs in \RRO\ behave like a canonical filling-controlled Mott transition as in the $3d$-electron Mott systems. Nevertheless, the limitation of such a simple framework become clear with increasing hole-doping level in the correlated metallic region. As described above, in the $4d$-electron systems, the correlated electrons renormalized by the multi-orbital effect are anticipated to dominate electronic and magnetic properties in a manner different from the simple case of the doped Mott insulator. Experimentally, Hund's metal regime has been less explored, in particular for pyrochlore compounds. In this study, we investigate the electronic states of the pyrochlore ruthenates systematically in a wide range of filling, which significantly varies the electron correlations induced by $U$ and $J_{\mathrm{H}}$. We successfully synthesized the mixed crystals of \PCRO\ in the whole composition range, $0 \leq x \leq 1$. Remarkably, on the Ca-rich side ($x \geq 0.5$), we find the emergence of ferromagnetism characteristic of Hund's metals. This marks the rare case of a ferromagnetic-metal pyrochlore. Based on the easily spin-polarized tendency of Hund's metal, we also clarify the large geometrical (topological) Hall effects due to the scalar spin chirality emerging from the $4f$(Pr)-$4d$(Ru) spin coupling.

Polycrystalline samples of \PCRO\ and \GCRO\ were prepared by using a cubic-anvil high-pressure apparatus; detailed procedures are described in Supplementary Material. Resistivity measurements were conducted by a standard 4-probe method, forming the contact with gold wire attached by silver paste. The magnetization was measured by a superconducting quantum interference device (SQUID) magnetometer. The optical reflectivity spectra were measured between 0.005 and 30 eV using a Fourier transform infrared spectrometer (0.05-0.5~eV), a grating monochromator (0.5-5~eV). For the Kramers-Kronig analysis to deduce the optical conductivity spectra, the room-temperature vacuum-ultra-violet reflectivity data above 5~eV measured at UVSOR (BL-7B), Institute for Molecular Science, were utilized.

\begin{figure}%[tbp]
\centering
\includegraphics[width=3.5in,keepaspectratio=true]{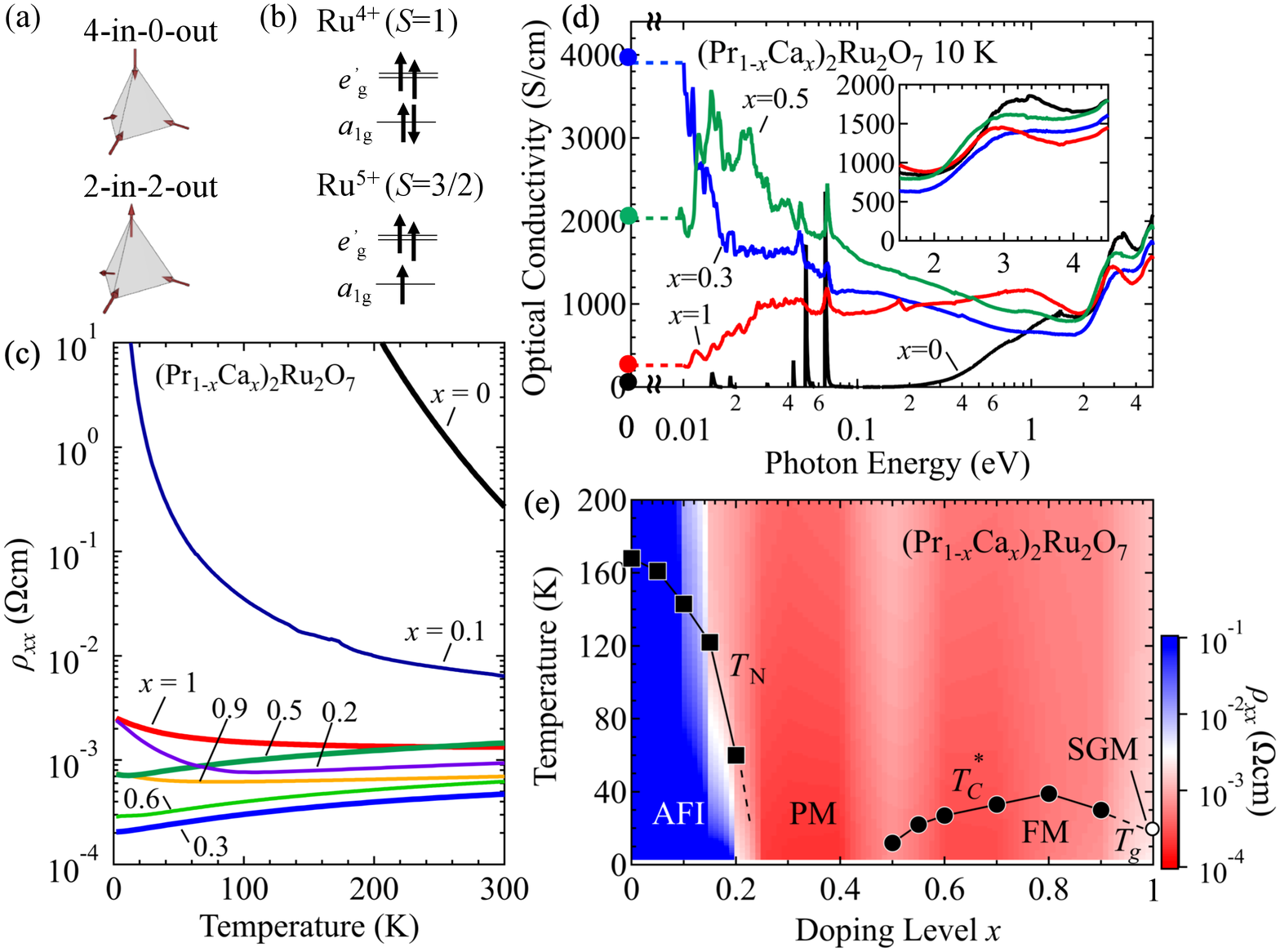}%,clip
\caption{\label{fig1} (a) All-in-all-out and 2-in-2-out magnetic structures. (b) $e'_{g}$ and $a_{1g}$ electronic levels and spin states of Ru$^{4+}$ and Ru$^{5+}$ in the ionic model. (c) Temperature dependence of longitudinal resistivity $\rho_{xx}$ for \PCRO. (d) Optical conductivity spectra of \PCRO\ ($x$ = 0, 0.3, 0.5 and 1) below the photon energy of 5 eV. (e) Contour plots of $\rho_{xx}$ on the doping level $x$ versus temperature plane. AFI, PM, FM and SGM represent antiferromagnetic insulator, paramagnetic metal, ferromagnetic metal and spin-glass metal, respectively. Black squares, black circles, and a white circle indicate the magnetic transition temperatures $T_\mathrm{N}$ (antiferromagnic), $T^{*}_{\mathrm{c}}$ (ferromagnetic), and $T_{\mathrm{g}}$ (spin glass), respectively.}
\end{figure}

Figure~1(d) shows the $x$ dependence of longitudinal resistivity $\rho_{xx}$ for \PCRO. The end compounds, $\mathrm{Pr}_2\mathrm{Ru}_2\mathrm{O}_7$ and $\mathrm{Ca}_2\mathrm{Ru}_2\mathrm{O}_7$, show insulating and bad-metallic behavior, respectively, being typical of the highly correlated electronic state. However, with carrier doping, namely nominal hole-doping into $\mathrm{Pr}_2\mathrm{Ru}_2\mathrm{O}_7$ or electron-doping to $\mathrm{Ca}_2\mathrm{Ru}_2\mathrm{O}_7$, $\rho_{xx}$ systematically decreases to show a good metallic behavior, as typically seen for $0.3 \leq x \leq 0.9$. The change of the electronic structure can be also inferred from the optical conductivity spectra for $x$ = 0, 0.3, 0.5 and 1 [Fig.~1(c)]; for $x$ = 0 ($\mathrm{Pr}_2\mathrm{Ru}_2\mathrm{O}_7$) a clear Mott-Hubbard gap of about 0.3 eV is observed, while for $x$ = 0.3 a clear Drude tail is discerned below 0.02 eV, followed by the broad spectrum in a higher photon energy region (0.02-1 eV). For $x$ = 1 ($\mathrm{Ca}_2\mathrm{Ru}_2\mathrm{O}_7$), the nearly $\omega$-flat conductivity spectrum dominates below 1 eV nearly down to 0.01 eV, corroborating the highly incoherent charge dynamics or bad-metal state and hence characterizing the incipient Mott insulator (the peaks at 0.01-0.07 eV are due to the optical phonons). Also characteristic of $\mathrm{Ca}_2\mathrm{Ru}_2\mathrm{O}_7$ is the red shift of the charge transfer excitation from occupied O-$2p$ to unoccupied (or upper Hubbard band like) Ru-$4d$ states, as discerned around 3 eV in the inset of Fig.~1(c). This evidences the relative upward shift of the O-$2p$ state due to the high-valence (5+) character of the Ru ions in $\mathrm{Ca}_2\mathrm{Ru}_2\mathrm{O}_7$.  

Figure~1(e) displays a contour plot of $\rho_{xx}$ together with the assignment of the phases in the plane of $x$ and temperature ($T$), where red and blue circles represent the magnetic transition temperatures determined from the anomaly in the temperature dependence of the magnetization. Three distinct phases are discernible in the contour plot. In the small $x$ region ($x < 0.3$) there persists the antiferromagnetic insulator phase, whose N\'{e}el temperature decreases with $x$. Then, the paramagnetic metallic phase appears for $0.3 \leq x < 0.5$, and above $x$ = 0.5 a ferromagnetically ordered metallic phase appears below 40 K while keeping a metallic value of $\rho_{xx}$ ($< 10^{-3}$~$\Omega\mathrm{cm}$), as later described in detail. Incidentally, just around $x$ = 0.5 a singular increment of resistivity is discerned in the whole temperature range (Fig.~1(c)), as exemplified in $\sigma_{xx} \sim \rho_{xx}^{-1}$ at 2 K shown in Fig.~4(b). The optical conductivity spectrum for $x$ = 0.5 [Fig.~1(d)] shows the metallic response down to lower energy, yet the pseudo-gap like or soft gap feature appears below 0.02~eV with vanishing Drude component. Though the origin of this insulating tendency around $x$ = 0.5 is unclear at the moment, one possibility is a formation of short-range fluctuating charge order such as an alternate nominal Ru$^{4+}$/Ru$^{5+}$ configuration on the frustrated pyrochlore lattice~\cite{Cheong1994}.

Next, we proceed to the magneto-transport properties in the paramagnetic metal regime. In Fig.~2(a), we show the magnetic field dependence of Hall conductivity $\sigma_{xy}$ of \PCRO\ for $x$ = 0.4 at various temperatures. The $\sigma_{xy}$ shows a linear field dependence at high temperatures, but with decreasing the temperature below 20 K, $\sigma_{xy}$ shows a highly field-nonlinear feature, i.e., the anomalous Hall effect (AHE) in a broad context. Because the $R$-$4f$ moments usually begin to polarize at such a low temperature, it is natural to consider that coupling between local $R$-$4f$ moments and conductive Ru-$4d$ electrons contributes to the observed AHE. To compare the behavior of $\sigma_{xy}$ among the \RRO\ compounds with magnetic $R$ ion with different magnetic anisotropies, $\sigma_{xy}$ of \GCRO\ for $x$ = 0.4 is shown in Fig.~2(b). Remarkably, the absolute value of $\sigma_{xy}$ at $B$ = 14 T is an order of magnitude larger for \PCRO\ than for \GCRO\ although the longitudinal conductivity $\sigma_{xx}$ values are both $> 10^3~\mathrm{S/cm}$; the $\rho_{xx}$ data of \GCRO\ are shown in Fig. S1 (Supplementary Material). Such a contrast will be attributed to the different $R$-$4f$ moments between \PCRO\ and \GCRO; the former being of Ising type with the magnetic anisotropy along $\langle 111 \rangle$ while the latter of Heisenberg type. In \PCRO, Ru-$4d$ electrons conducting on a non-collinear spin texture composed of Pr-$4f$ moments with finite scalar spin chirality (SSC) experiences an emergent magnetic field derived from the real-space Berry phase~\cite{Nagaosa2010}. The resultant transverse response of the electrons is called topological or geometrical Hall effect (GHE) to be distinguished from the intrinsic (Karplus-Luttinger type) AHE due to SOC; the GHE is now widely observed in frustrated or chiral magnets~\cite{Taguchi2001,Lee2009,Machida2009,Wang2019,Kurumaji2019}. 

\begin{figure}[tbp]
\centering
\includegraphics[width=3.5in,keepaspectratio=true,clip]{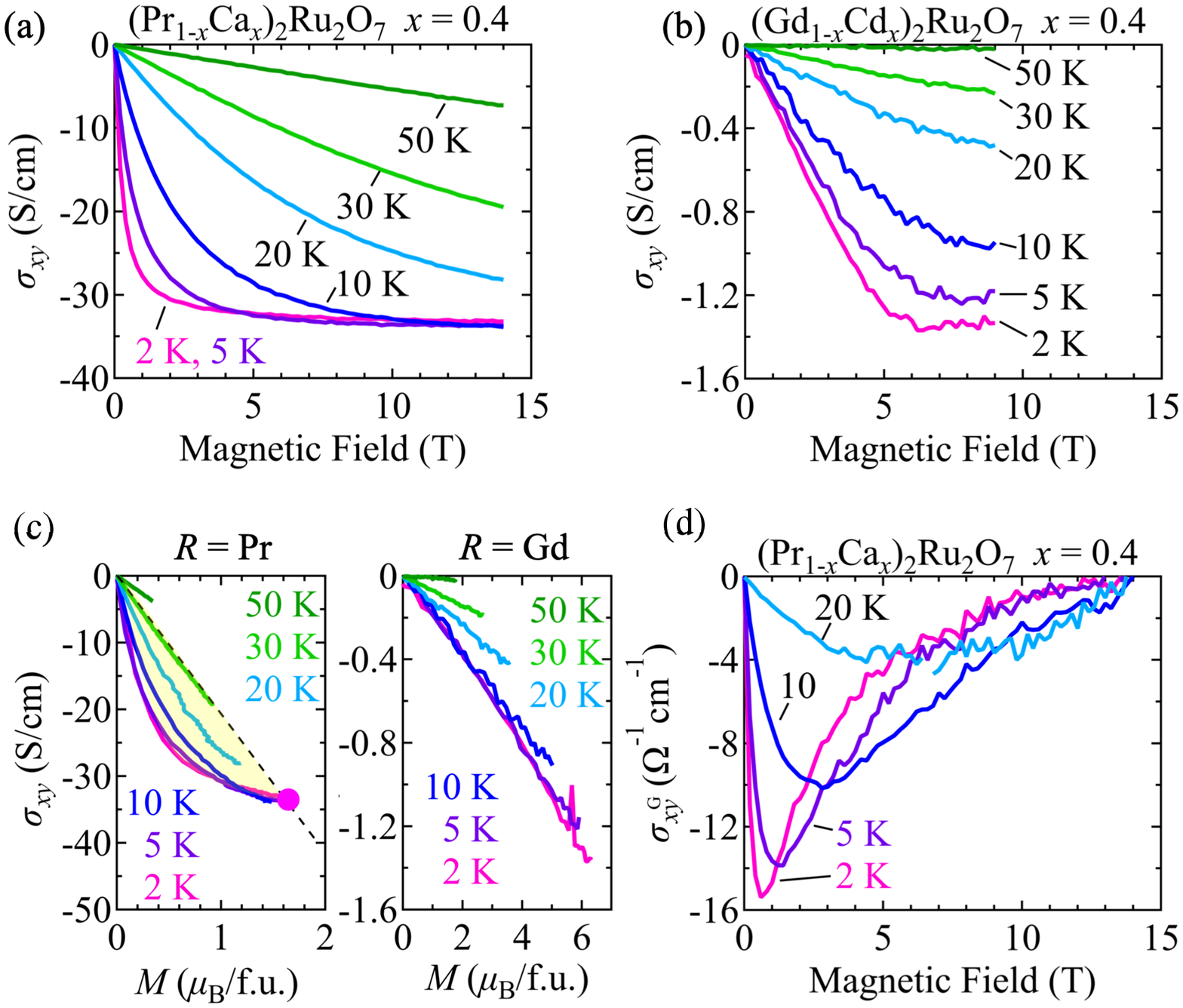}
\caption{\label{fig2} Magnetic field dependence of Hall conductivity $\sigma_{xy}$ in the paramagnetic phases of (a) \PCRO\ ($x$ = 0.4) and (b) \GCRO\ ($x$ = 0.4) at various temperatures. (c) $\sigma_{xy}$ plotted as a function of magnetization $M$ for \PCRO\ of $x = 0.4$ (left) and \GCRO\ of $x=0.4$ (right). Yellow hatched region in the left panel highlights the geometrical (topological) Hall term $\sigma^{\mathrm{G}}_{yx}$ at 2 K. (d) Magnetic field dependence of $\sigma^{\mathrm{G}}_{yx}$ for \PCRO\ ($x$ = 0.4). }
\end{figure}

To analyze the contribution of the $f$-$d$ coupling and the resultant GHE, we use the phenomenological relation for the AHE~\cite{Nagaosa2010},
\begin{equation}
\rho_{yx} =  R_0 H + S_{\mathrm{A}}\rho_{xx}^2 M + \rho^{\mathrm{G}}_{yx},
\end{equation}
where $R_0$ and $S_{\mathrm{A}}$ are the ordinary and anomalous Hall coefficients, respectively. The second term is the $M$-linear term mediated by the SOC, including the intrinsic (Karplus-Luttinger) mechanism, and the third term is the geometrical term mediated by the SSC, the source of the GHE in our case. Figure~2(c) plots the $M$ dependence of Hall conductivity $\sigma_{xy} = \rho_{yx}/(\rho^2_{xx} +\rho^2_{yx})$ for \PCRO\ and \GCRO\ ($x = 0.4$). Here we assume that the ordinary Hall coefficient is negligible ($|R_0| \sim 10^{-3}$~$\mathrm{cm^3/C}$, a typical value of metallic pyrochlore oxides~\cite{Yoshii2000,Machida2007}). For \PCRO, $\sigma_{xy}$ is not proportional to the magnetization $M$, indicating that the SSC originating from the Ising-type Pr-$4f$ moments causes the GHE. In contrast, for \GCRO, where the Heisenberg-type Gd-$4f$ moments produce no static component of SSC, $\sigma_{xy}$ is almost proportional to $M$, i.e., dominated by the conventional AHE.

To extract the contribution of the GHE, we first estimate the $M$-linear AHE. At low temperatures below 20 K, $\sigma_{xy}$ rapidly saturates with increasing magnetic field above 10 T [Fig.~2(a)], implying the vanishing SSC in the fully field-aligned collinear Ru spins, as observed in $\sigma_{xy}$ in the archetypal pyrochlore magnet $\mathrm{Nd}_2\mathrm{Mo}_2\mathrm{O}_7$ with the SSC~\cite{Taguchi2001}. Thus $S_A$ of \PCRO\ for $x = 0.4$ can be deduced from the slope of the line connecting the $\sigma_{xy}$ data points at $M$=0 ($B$ = 0 T) and at the saturated $M$ (e.g. $B$=14 T) in the $M$-$\sigma_{xy}$ plot (a dashed line in the left panel of Fig.~2(c)). Note that the 30 K data almost coincide with this relation, meaning the SSC fading at and above this temperature. After subtracting this $M$-linear AHE term, the GHE conductivity term $\sigma^{\mathrm{G}}_{yx}$ is obtained as shown in Fig.~2(d). At 2 K, $\sigma^{\mathrm{G}}_{xy}$ sharply arises with the magnetic field, takes a peak, and gradually vanishes. This is a consequence of the competition between the Zeeman energy favoring the collinear Ru spin alignment and the magnetic anisotropic energy forming the SSC which stems from the exchange interaction between Ru spins and Ising Pr-$4f$ moments. With increasing the temperature, the peak value decreases and its magnetic-field value  shifts to a higher value, as expected from the thermal fluctuation of the Pr-$4f$ moments. 

Contrary to the hole-doped paramagnetic metal regime ($0.3 \leq x < 0.5$), in the electron-doped regime of \PCRO\ ($0.5 \leq x < 1$), the interaction among Ru-$4d$ electrons causes the ferromagnetic correlation even at zero field, whereas their coupling to the Pr-$4f$ moments causes the both $4d$- and $4f$-moment-coupled magnetic order. In this regime, the magnetic transition is discerned from a bifurcation between the field-cooling and zero-field-cooling curves in the temperature-magnetization measurement. The transition temperature $T^*_c$ estimated from the magnetization $M$ is plotted in Fig.~1(d) with black circles.
%A similar transition can be seen in AC susceptibility $\chi_\mathrm{AC}$ for some compounds of \PCRO, although the anomaly temperatures in $\chi_\mathrm{AC}$ are different from those observed in $M$ (shown in SM). Such a difference is typical of the diluted ferromagnetic phase known as the Griffiths phase~\cite{Deisenhofer2005}. 
%
When the temperature is low enough below $T^*_c$, the field dependence of $M$ [Figs.~3(c), (d)] and $\sigma_{xy}$ [Figs.~3(f), (g)] of \PCRO\ ($x$ = 0.7, 0.9) commonly show hysteresis curves, indicating the existence of ferromagnetic domains. The spontaneous magnetization $M_s$ and the Hall conductivity at 0 T and 4 T are plotted in Figs.~4(a) and (c), indicating that the ferromagnetic-paramagnetic boundary is located around $x$ = 0.5. On the other hand, for $x$ = 1 ($\mathrm{Ca}_2\mathrm{Ru}_2\mathrm{O}_7$), where the spin-glass transition has been observed at $T_\mathrm{N}$ = 25 K in previous studies~\cite{Munenaka2006}, no clear hysteresis is observed even at the lowest temperature (2 K), as shown in Figs.~3(e) and (h). For all values of $x$ $M$ does not saturate even at $B$ = 7 T, indicating an antiferromagnetic coupling between Pr-$4f$ moment and Ru-$4d$ spin, as observed in pyrochlore ferromagnet $\mathrm{Nd}_2\mathrm{Mo}_2\mathrm{O}_7$~\cite{Yasui2003}.
Figure~4(a) plots the Ru-moment part of the spontaneous magnetization ($M^\mathrm{Ru}_s = M_s - (- M^\mathrm{Pr})$) as green squares, as analyzed by assuming the antiferromagnetic $f$-$d$ coupling.
Here, the Pr-moment part of the magnetization is estimated by $M^\mathrm{Pr} = M^\mathrm{Pr} _{\mathrm{sat}} (1 - x)$, where $M^\mathrm{Pr}_{\mathrm{sat}}$ (= 2.5~$\mu_\mathrm{B}$/Pr) is the saturation magnetism from Pr$^{3+}$ moments in a polycrystalline form, which is approximated by $M$ at $B$ = 7 T for the nondoped $\mathrm{Pr}_2\mathrm{Ru}_2\mathrm{O}_7$. The value of $M^\mathrm{Ru}_s$ appears considerably lower than the fully polarized value, $\sim (4 + 2x)~\mu_\mathrm{B}$/f.u. in the ionic picture, due mostly to the itinerant character of the Ru electrons and due partly to the size effect of the crystal grains. We note that the above assumption of the robustness of the Ru ferromagnetism, which is antiferromagnetically coupled with the full Pr magnetization $M^{\mathrm{Pr}}_{\mathrm{sat}}$, becomes unjustified in the vicinity of the paramagnetic phase boundary ($x \sim 0.5$), where $M^\mathrm{Ru}_s$ will be overestimated.

\begin{figure}[tbp]
\centering
\includegraphics[width=3.5in,keepaspectratio=true]{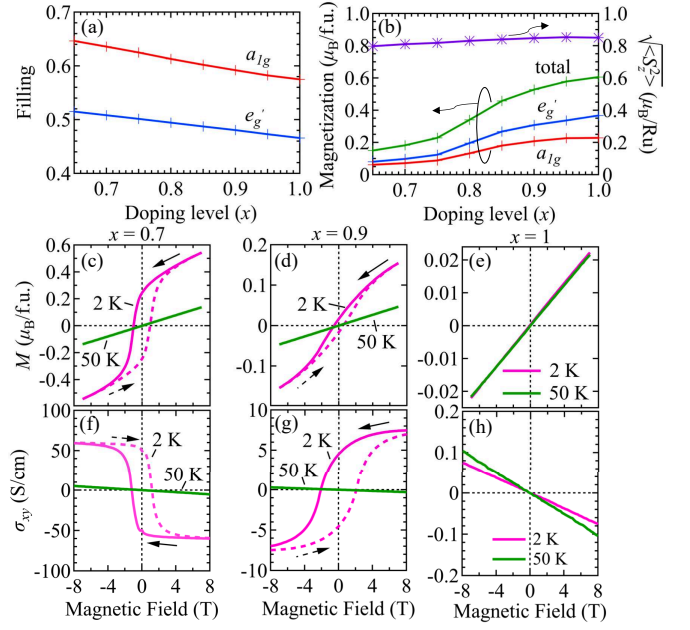}
\caption{\label{fig3} DFT+DMFT results for (a) the orbital occupancy and (b) magnetization plotted against $x$. Red and blue crosses represent the results for $a_{1g}$ and $e'_g$ orbitals, respectively. Purple marks in (b) represent the expectation value of local moment $\sqrt{\langle S_z^2 \rangle}$. Magnetic field dependence of (c-e) magnetization $M$ and (f-h) Hall conductivity $\sigma_{xy}$ for \PCRO\ ($x$ = 0.7, 0.9, 1) at temperatures, 2 K and 50 K, below and above $T^*_{\mathrm{c}}$, respectively. }
\end{figure}
\begin{figure}[tbp]
\centering
\includegraphics[width=3in,keepaspectratio=true]{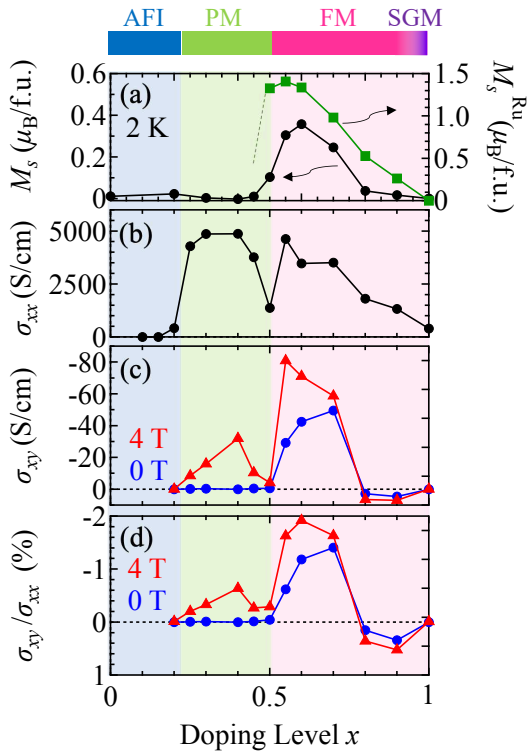}
\caption{\label{fig4} Doping level $x$ dependence of (a) spontaneous magnetization $M_s$, (b) longitudinal conductivity $\sigma_{xx}$, (c) Hall conductivity $\sigma_{xy}$ and (d) Hall angle $\sigma_{xy}$/$\sigma_{xx}$ at 2 K. Green squares in (a) show the estimated value of spontaneous magnetization for the Ru-$4d$ moment, where the antiferromagnetic $f$-$d$ coupling such as $M_s = M^\mathrm{Ru}_s - M^\mathrm{Pr}$ is assumed. Shaded colors and color bar highlight the corresponding electronic and magnetic phases.}
\end{figure}

To understand the mechanism of ferromagnetism in the electron-doped regime, i.e., the higher-$x$ region near $\mathrm{Ca}_2\mathrm{Ru}_2\mathrm{O}_7$, we theoretically investigate the electronic structure of \PCRO. The calculation is based on density functional theory combined with the dynamical mean-field theory (DMFT). We find that the DMFT indeed has a ferromagnetic solution for $0.65 < x \leq 1$, where $J_{\mathrm{H}}$ reduces the quasiparticle weight in line with the Hund's-metal picture (see Supplementary Material for details). Figures~3(a) and 3(b) plot the calculated orbital filling and $M$ against the doping level ($x$). The filling of the $a_{1g}$ and $e'_g$ orbitals similarly increases with electron doping $(1-x)$ to $\mathrm{Ca}_2\mathrm{Ru}_2\mathrm{O}_7$, while the magnetization systematically decreases. Such behaviors suggest that both $a_{1g}$ and $e'_g$ orbitals have an itinerant character, unlike a simple double-exchange mechanism which accounts the distinct roles of the localized $a_{1g}$ and itinerant $e'_g$ electrons. Nevertheless, Hund's coupling $J_{\mathrm{H}}$, which less contributes to the Stoner mechanism for the itinerant magnetism, plays a crucial role in the present ferromagnetism~\cite{Sakai2007}.
This is substantiated by the result that the ferromagnetism disappears when we artificially set $J_{\mathrm{H}}=0$. This would also explain the suppression of $M$ with decreasing $x$ since $J_{\mathrm{H}}$ would become less effective from Ru$^{5+}$ ($S=3/2$) to Ru$^{4+}$ ($S=1$). The calculated $M$ values (0.2-0.6~$\mu_\mathrm{B}$/f.u.) are comparable with the experimentally observed values (see Fig.~4(a)) apart from the above-mentioned possible overestimate of the experimental values at $x$ = 0.5-0.6, close to the paramagnetic-phase boundary. Despite this small $M$, the calculated local moment $\sqrt{\langle S_z^2 \rangle}$ is rather large ($\sim 0.8\ \mu_\mathrm{B}$/Ru). This would be consistent with the picture that conduction electrons align large local moments fluctuating dynamically. However, the DMFT does not reproduce the experimental $x$ dependence of $M$. In particular, near $x$ = 1, as evidenced by the experimental observation of spin-glass phase, the competition between the ferromagnetic and antiferromagnetic interactions as well as the magnetic frustration on the pyrochlore lattice would be necessary to explain this discrepancy.

Finally, we sum up the whole $x$ dependence of the magnetotransport properties. In the analysis of $\sigma_{xy}$ in the paramagnetic metal region of \PCRO, we show that the Ising-anisotropic moments of Pr induces the SSC of the Ru spins via the exchange coupling between the Pr-$4f$ moments and the Ru-$4d$ spins, resulting in the GHE. In \PCRO, the magnitude of the GHE should be proportional to the possibility that the Pr-$4f$ moments form the triad, i.e., $(1 - x)^3$~\cite{Ueda2012}. However, $\sigma_{xy}$ plotted in Fig.~4(c) deviates from the $(1 - x)^3$ dependence, indicating that the intrinsic AHE of SOC origin is not negligible in the magneto-transport of \PCRO. The intrinsic AHE, relating to the $k$-space Berry curvature, is typically enhanced when the Fermi energy approaches the anti-crossing points or Weyl points in the band structure ~\cite{Fang2003,Shekhar2018}. In our case, $|\sigma_{xy}|$ increases systematically with increasing $x$ until around $x$ = 0.6, where the Hall angle lifts up to 2 \%, and the sign change occurs between $x$ = 0.7 and $x$ = 0.8 [Fig.~4(d)]. This $x$ dependence of $\sigma_{xy}$ suggests that the Fermi level may cross one of the Weyl nodes involving a large Berry curvature, along with the carrier doping. 

In summary, we investigate the electronic and magnetic properties in the mixed crystals of pyrochlore \PCRO, in which the band filling is fully controlled from the $x$ = 0 antiferromagnetic Mott insulator to the $x$ = 1 spin-glass like correlated bad metal. We find that, as $x$ increases, the system undergoes the transition from the antiferromagnetic Mott insulator to the paramagnetic metal, where the $f$-$d$ coupling between the anisotropic Pr-$4f$ moments and conductive Ru-$4d$ electrons dominates the magneto-transport properties such as the geometrical Hall effect. With further increasing $x$ above 0.5 or doping electrons to $\mathrm{Ca}_2\mathrm{Ru}_2\mathrm{O}_7$, the compounds exhibit metallic ferromagnetism, where the $4d$ electrons are strongly correlated by the multiorbital effect, being typical of Hund's metals. These phase transitions are accompanied by the large anomalous Hall effect with the non-trivial $x$ dependence, highlighting the easily spin-polarized correlated metallic state affected by Hund's coupling. 

We thank Y. Kaneko and A. Kikkawa for helpful advice about crystal growth. This work was supported by the Japan Society for the Promotion of Science (KAKENHI; Grants No. 19K14647) from the MEXT, and by CREST (Grant No. JPMJCR16F1 and JPMJCR1874) from Japan Science and Technology Agency.
%K.U. and Y.T. conceptualized the work; R.K. conducted the experiments.

%\nocite{*}
%
%\bibliographystyle{myapsrev4-1}%myapsrev4-2
%\bibliography{bibfile2.bib}
%

\end{document}